\documentclass[12pt]{iopart}%
\usepackage[dvips]{graphics}
\usepackage{graphicx}
\usepackage{bm}
\usepackage{amsfonts}
\usepackage{amssymb}
\usepackage{cite}%
\providecommand{\U}[1]{\protect\rule{.1in}{.1in}}
\begin{document}

\title{Thermoelectric response of spin polarization in Rashba spintronic systems}
\author{Cong Xiao$^{1,2}$, Dingping Li$^{1,2}$, and Zhongshui Ma$^{1,2}$}
\address{$^{1}$School of Physics, Peking University, Beijing 100871, China}
\address{$^{2}$Collaborative Innovation Center of Quantum Matter, Beijing, 100871, China}

\begin{abstract}
Motivated by recent discovery of strongly spin-orbit coupled two-dimensional
(2D) electron gas near the surface of Rashba semiconductors BiTeX (X=Cl, Br,
I), we calculate thermoelectric responses of spin polarization in 2D Rashba
model. By self-consistently determining the energy-and subband-dependent
transport time, we present an exact solution of the linearized Boltzmann
equation for elastic scattering. Using this solution, we find a non-Edelstein
electric-field induced spin polarization which is linear in the Fermi energy
$E_{F}$, when $E_{F}$ lies below the band crossing point. The spin
polarization efficiency, which is the electric-field induced spin polarization
divided by the driven electric current, increases for smaller $E_{F}$. It is
shown that, as a function of $E_{F}$, the temperature-gradient induced spin
polarization continuously increases to a saturation value when $E_{F}$ downs
below the band crossing point. As the temperature tends to zero, the
temperature-gradient induced spin polarization vanishes.

\vspace{3mm}
\end{abstract}

\section{Introduction}

Two dimensional (2D) electron systems with spin-orbit coupling (SOC) show a
great deal of fascinating transport phenomena due to the mixing of the spin
and orbital degrees of freedom, providing the possibility of realizing
all-electrical and all-thermal spin control in semiconductor structures. These
are the main topics of the rapid developing research fields of
spintronics\cite{Zutic2004} and spin-caloritronics\cite{Bauer2012}. In the electrical spin control, the generation of a spin current
and a nonequilibrium spin polarization transverse to an applied electric field
without external magnetic field are remarkable, known as the spin Hall
effect\cite{Sinova2004,Hirsch1999} and electric-field induced spin
polarization\cite{Edelstein1990,Mishchenko2004,Gorini2008}, respectively. Their thermal counterparts toward all-thermal spin
control, i.e., the spin Nernst effect\cite{Ma2010,Akera2013,Borge2013,Iglesias2014,Tauber2012} and
temperature-gradient induced spin polarization\cite{Wang2010,Dyrdal2013,Toelle2014}, have also attracted more and more interests recently.

The 2D electron system (2DES) with Rashba SOC has been one of the most widely
used models to investigate aforementioned effects\cite{Ma2010,Borge2013,Wang2010,Dyrdal2013}. In the 2D Rashba model, two
bands cross at zero energy, one of them is always positive and the other one
possesses a band valley regime below the band crossing point as shown in Fig.
1. In this valley regime the dispersion curve is not monotonic in momentum
space. This regime possesses nontrivial topology of the constant energy
surfaces (or Fermi surfaces)\cite{Cappelluti2007}, which leads to
some exciting theoretical predictions, e.g., the enhanced superconducting
critical temperature\cite{Cappelluti2007}, the non-Dyakonov-Perel
spin relaxation behavior\cite{Grimaldi2005} and the significantly enhanced
room-temperature thermoelectric figure of merit\cite{Wu2014b}. There
have been a few theoretical studies \cite{Grimaldi2005,Grimaldi2006,Tsutsui2012,Dyrdal2013,Wu2014b} on the
transport properties when the Fermi energy is in or near the band valley
regime. However, in Rashba systems formed in conventional narrow-gap
semiconductor heterostructures\cite{Nitta1997}, the Rashba spin
splitting energy is so small that the band valley structure can not survive
the weak disorder broadening and thermal smearing even at very low
temperatures. In these systems the Fermi energy usually lies quite above the
band crossing point, therefore the band valley is irrelevant to transport.

Recently experimental progress has been made by the discovery of giant bulk
and surface Rashba SOC effects in V-VI-VII polar semiconductors BiTeX, (X=Cl,
Br, I)\cite{Eremeev,Landolt,Crepaldi2012}. In these
noncentrosymmetric semiconductors, first-principles calculations and ARPES
measurements have clearly demonstrated the existence of 2DES confined near the
surface with giant Rashba energy as large as about $10^{2}meV$\cite{Sakano2013,Rusinov2013}. In such 2DES, the investigation of electrical
and thermal spin control is of significance due to the giant SOC which is
promising for spintronics and spin-caloritronics applications. While, when the
electron-impurity scattering dominates, for the case that Fermi energies lie below or in the vicinity of
the band crossing point, the relaxation time approximation (RTA)
used in previous theoretical works on the nonequilibrium spin polarization\cite{Tsutsui2012,Dyrdal2013}
may not work well due to the giant Rashba SOC. This motivates us to systematically investigate the
thermoelectric response of spin polarization in 2DES with giant Rashba SOC,
focusing on the consequences of different Fermi surface topologies between the
two sides of the band crossing point.

In this paper, we employ the semiclassical Boltzmann equation to calculate the
spin polarization induced by electric field and temperature gradient. We focus
on the 2D Rashba model at low temperatures where the static impurity
scattering dominates. Our calculation is based on an exact transport time
solution of the Boltzmann equation in the Born approximation, different from
the widely used modified RTA and constant RTA schemes\cite{Ziman1972}. We
show that the electric-field induced spin polarization (EISP) as a function of
the Fermi energy $E_{F}$ behaves differently between the two sides of the band
crossing point $E_{F}=0$. A linear dependence of EISP on $E_{F}$ is obtained
for $E_{F}<0$, differing from the Edelstein result\cite{Edelstein1990} for
$E_{F}\geq0$. The spin polarization efficiency, defined as the ratio between
the EISP and the driven electric-current, increases for lower $E_{F}$. The
temperature-gradient induced spin polarization (TISP) is calculated, and its
dependence on the Fermi energy, changing from large positive values to be
quite below the band crossing point, is continuous and monotonic. It is also
shown that the temperature-gradient induced spin polarization tends to zero at
vanishing temperatures.

\section{Semiclassical Boltzmann descriptions of thermoelectric spin responses
in Rashba 2DES}

\subsection{Basic solutions for the 2D Rashba model}

We study the 2D Rashba model with spin independent disorder%
\begin{equation}
H=\frac{\mathbf{p}^{2}}{2m}+\frac{\alpha}{\hbar}\mathbf{\sigma}\cdot\left(
\mathbf{p}\times\mathbf{\hat{z}}\right)  +V\left(  \mathbf{r}\right)  ,
\end{equation}
where $V\left(  \mathbf{r}\right)  =\sum_{i}V_{i}\delta\left(  \mathbf{r}%
-\mathbf{R}_{i}\right)  $ is the disorder potential produced by randomly
distributed $\delta$-scatters at $\mathbf{R}_{i}$ and is assumed to be
standard white-noise disorder: $\left\langle \left\vert V_{\mathbf{k}^{\prime
}\mathbf{k}}\right\vert ^{2}\right\rangle _{dis}=n_{im}V_{0}^{2}$. Here
$n_{im}$ is the impurity concentration, $V_{\mathbf{k}^{\prime}\mathbf{k}}$
the spin-independent part of the disorder matrix element and $\left\langle
..\right\rangle _{dis}$ the disorder average. $m$ is the in-plane effective
mass of the conduction electron, $\mathbf{p=\hbar k}$ the momentum,
$\mathbf{\sigma=}\left(  \sigma_{x},\sigma_{y},\sigma_{z}\right)  $ are the
Pauli matrices, $\alpha$ the Rashba coefficient.
Eigenenergies of the pure system are $E_{\lambda k}=\frac{\hbar^{2}k^{2}}%
{2m}+\lambda\alpha k$, with inner eigenstates $|u_{\lambda\mathbf{k}}%
\rangle=\frac{1}{\sqrt{2}}\left[  1,-i\lambda\exp\left(  i\phi\right)
\right]  ^{T}$, where $\lambda=\pm$ and $\tan\phi=k_{y}/k_{x}$. The wave
number at a given energy $E>0$ in the $\lambda$ band is given as $k_{\lambda
}\left(  E\right)  =-\lambda k_{R}+\frac{1}{\alpha}\sqrt{E_{R}^{2}+2E_{R}E}$
(see Fig. 1), where we define the Rashba energy $E_{R}=m\left(  \frac{\alpha
}{\hbar}\right)  ^{2}$ and $k_{R}=\frac{E_{R}}{\alpha}$. The density of state
(DOS) at a given $E\geq0$ is given by $N_{>}\left(  E\right)  =\sum_{\lambda
}N_{\lambda}\left(  E\right)  $ where
\begin{equation}
N_{\lambda}\left(  E\right)  =N_{0}\frac{k_{\lambda}\left(  E\right)
}{k_{\lambda}\left(  E\right)  +\lambda k_{R}}. \label{DOS1}%
\end{equation}
Here $N_{0}=\frac{m}{2\pi\hbar^{2}}$ is the DOS of 2D spin polarized parabolic spectrum.
For $E>0$, the group velocity and intraband spin matrix element are given by
\begin{equation}
\mathbf{v}\left(  E,\lambda,\phi\right)  =\frac{N_{0}}{N_{\lambda}\left(
E\right)  }\frac{\hbar\mathbf{k}_{\lambda}\left(  E\right)  }{m}
\label{velocity1}%
\end{equation}
and%
\begin{equation}
\langle u_{\lambda\mathbf{k}_{\lambda}\left(  E\right)  }|\mathbf{\sigma
}|u_{\lambda\mathbf{k}_{\lambda}\left(  E\right)  }\rangle=\lambda\left(
\sin\phi\mathbf{\hat{x}-}\cos\phi\mathbf{\hat{y}}\right)  ,
\end{equation}
respectively.

\begin{figure}[ptbh]
\begin{indented}
\item[]\includegraphics[width=0.36\textwidth]{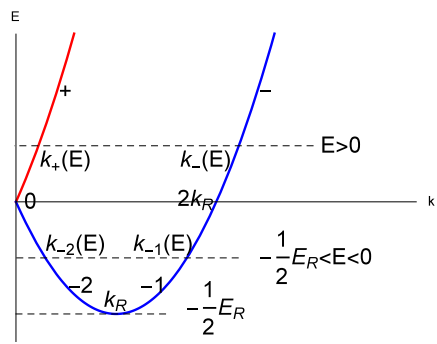}
\end{indented}
\caption{Band structure of the 2D Rashba system. The energy of the band
crossing point is chosen to be zero. The wave number and energy of the bottom
of the dispersion curve is $k_{R}$ and $-\frac{1}{2}E_{R}$, respectively.
Corresponding to a given energy $E\geq0$, the wave number in $\pm$ band is
denoted by $k_{\pm}\left(  E\right)  $. For $-\frac{1}{2}E_{R}<E\leq0$, there
are two monotonic regimes on $E-k$ curve: the one from $k=0$ to $k_{R}$ is
marked by the branch $-2$, whereas the other from $k=k_{R}$ to $2k_{R}$ marked
by branch $-1$. The wave number $k_{-\nu}\left(  E\right)  $ represents the
wave number in the $-\nu$ branch at given $E$, where $\nu=1,2$.}%
\label{fig1}%
\end{figure}The direction of the group velocity is the same as the
corresponding momentum due to the isotropic band structure and the monotonic
$E-k$ curve when $E\geq0$. The directions of spin in the two bands at the same
polar angle $\phi$ are opposite to each other.

The lower band has a valley centered at $k_{R}$, and the DOS has a
one-dimensional (1D) character in the $E_{-}\left(  k_{R}\right)  \leq E<0$
regime\cite{Cappelluti2007} with $E_{-}\left(  k_{R}\right)
=-\frac{1}{2}E_{R}$\ the energy of the bottom of dispersion curve. For
$E_{-}\left(  k_{R}\right)  <E<0$ there are two wave numbers $k_{-2}\left(
E\right)  <k_{R}<k_{-1}\left(  E\right)  $ with $k_{-\nu}\left(  E\right)
=k_{R}+\left(  -1\right)  ^{\nu-1}\frac{1}{\alpha}\sqrt{E_{R}^{2}+2EE_{R}}$,
where $-\nu=-1,-2$ denotes the two monotonic branches in this energy regime
(see Fig. 1). The DOS $N_{<}\left(  E\right)  $\ in the band valley regime is
given by $N_{<}\left(  E\right)  =\sum_{\nu=1}^{2}N_{-\nu}\left(  E\right)  $
where%
\begin{equation}
N_{-\nu}\left(  E\right)  =N_{0}\frac{k_{-\nu}\left(  E\right)  }{\left\vert
k_{-\nu}\left(  E\right)  -k_{R}\right\vert }. \label{DOS2}%
\end{equation}
For\textbf{\ }$E_{-}\left(  k_{R}\right)  <E<0$, one obtains%
\begin{equation}
\langle u_{-\mathbf{k}_{-\nu}\left(  E\right)  }|\mathbf{\sigma}%
|u_{-\mathbf{k}_{-\nu}\left(  E\right)  }\rangle=-\sin\phi\mathbf{\hat{x}%
+}\cos\phi\mathbf{\hat{y}}%
\end{equation}
and%
\begin{equation}
\mathbf{v}\left(  E,-\nu,\phi\right)  =\left(  -1\right)  ^{\nu-1}\frac{N_{0}%
}{N_{-\nu}\left(  E\right)  }\frac{\hbar\mathbf{k}_{-\nu}\left(  E\right)
}{m}. \label{velocity2}%
\end{equation}
The direction of the group velocity is parallel (anti-parallel) to the
corresponding momentum in $-\nu=-1$ $\left(  -2\right)  $ branch,
respectively. This is caused by the non-monotonic $E-k$ curve in the band
valley regime. The directions of spin in the two monotonic branches at the
same $\phi$ are the same. These characters show different spin-and group velocity-textures of
constant-energy circles between the band valley regime and the $E>0$ regime.

Exactly at the band crossing point $\left(  E,k\right)  =\left(  0,0\right)
$, the eigenstate as well as the group velocity and spin matrix element are
not well-defined because the polar angle $\phi$ is arbitrary. However, this
does not bring any influence on physical quantities since the DOS at this
point is zero, as shown in Eqs. (\ref{DOS1}) and (\ref{DOS2}). As for $\left(
E,k\right)  =\left(  0,2k_{R}\right)  $, the group velocity and spin matrix
element are both well-defined and continuous.

\subsection{Basic formulas of nonequilibrium spin polarization and Boltzmann equation}

The out-of-equilibrium spin density response to external fields can be
obtained in the semiclassical version of linear response theory, which can be
decomposed into the intrinsic and extrinsic parts%
\begin{equation}
\left\langle \mathbf{\hat{\sigma}}\right\rangle \equiv\left\langle
\mathbf{\hat{\sigma}}\right\rangle _{int}+\left\langle \mathbf{\hat{\sigma}%
}\right\rangle _{ext},
\end{equation}
where%
\begin{eqnarray}
\left\langle \mathbf{\hat{\sigma}}\right\rangle _{int}  &  =\sum_{l}f_{l}%
^{0}2{Re}\langle\psi_{l}^{\left(  0\right)  }|\mathbf{\sigma}|\delta\psi
_{l}\rangle,\label{SP1}\\
\left\langle \mathbf{\hat{\sigma}}\right\rangle _{ext}  &  =\sum_{l}%
g_{l}\langle\psi_{l}^{\left(  0\right)  }|\mathbf{\sigma}|\psi_{l}^{\left(
0\right)  }\rangle.\nonumber
\end{eqnarray}
Here $f_{l}^{0}$ is the equilibrium Fermi-Dirac distribution function (DF) and
$g_{l}$ denotes the out-of-equilibrium change of DF, $l=\left(  \lambda
,\mathbf{k}\right)  $ is the eigenstate index denoting the band and momentum.
$|\psi_{l}^{\left(  0\right)  }\rangle$ is the eigenstate of the disorder-free
Hamiltonian in the absence of external fields, $|\delta\psi_{l}\rangle$
describes the virtual interband transition induced by the weak external
fields. \

$\left\langle \mathbf{\hat{\sigma}}\right\rangle _{int}$ originates from the intrinsic mechanism based solely on the
spin-orbit coupled band structure. It is not difficult to verify that
$\left\langle \mathbf{\hat{\sigma}}\right\rangle _{int}=0$ for the present
model, so one only needs to analyze the extrinsic spin density response
$\left\langle \mathbf{\hat{\sigma}}\right\rangle _{ext}$ which depends on the
existence of disorder via $g_{l}$.

$g_{l}$ can be calculated by the semiclassical Boltzmann equation  in the
presence of an uniform weak electric field and small gradients of chemical potential and temperature in
nonequilibrium steady states. Here we consider low temperatures where the
static impurity scattering dominates the electron relaxation. The Boltzmann equation reads
\begin{equation}
\mathbf{F}_{l}\cdot\mathbf{v}_{l}\frac{\partial f^{0}}{\partial E_{l}}%
=-\sum_{l\prime}w_{l\prime,l}\left[  g_{l}-g_{l\prime}\right]  , \label{SBE}%
\end{equation}
where the generalized force acting on the state $l$ is $\mathbf{F}_{l}%
=-\frac{E_{l}-\mu}{T}\mathbf{\nabla}T-\mathbf{\nabla}\mu+e\mathbf{E}$ with
$\mathbf{E}$, $\mu$, $T$ being the electric field, chemical potential and
absolute temperature, respectively. $w_{l\prime,l}$ is the transition rate
from state $l^{\prime}$ to $l$, which can be determined by the golden rule
in the quantum mechanical scattering theory. In the present
system without anomalous Hall effect, the lowest order Born approximation is sufficient:%
\begin{equation}
w_{l\prime,l}=\frac{1}{\tau_{0}N_{0}}\left\vert \langle u_{l\prime}%
|u_{l}\rangle\right\vert ^{2}\delta\left(  E_{l}-E_{l\prime}\right)  ,
\label{rate}%
\end{equation}
with $\tau_{0}=\left(  \frac{2\pi n_{im}V_{0}^{2}N_{0}}{\hbar}\right)  ^{-1}$.
When $E>0$, the intraband and interband elastic scattering can be represented
by $\omega_{\lambda^{\prime},\lambda}^{\phi^{\prime},\phi}\left(
E=E_{l}\right)  =\int dE_{l^{\prime}}w_{l\prime,l}$:
\begin{equation}
\omega_{\lambda^{\prime},\lambda}^{\phi^{\prime},\phi}\left(  E\right)
=\frac{1}{\tau_{0}N_{0}}\frac{1}{2}\left[  1+\lambda\lambda^{\prime}%
\cos\left(  \phi^{\prime}-\phi\right)  \right]  .
\end{equation}
Whereas for $E_{-}\left(  k_{R}\right)  <E<0$ we introduce $\omega
_{-\nu^{\prime},-\nu}^{\phi^{\prime},\phi}\left(  E=E_{l}\right)  =\int
dE_{l^{\prime}}w_{l\prime,l}$ to represent the intra-branch and inter-branch
scattering:%
\begin{equation}
\omega_{-\nu^{\prime},-\nu}^{\phi^{\prime},\phi}\left(  E\right)  =\frac
{1}{\tau_{0}N_{0}}\frac{1}{2}\left[  1+\cos\left(  \phi^{\prime}-\phi\right)
\right]  .
\end{equation}

\section{The exact solution of the Boltzmann equation}

In this section we analytically solve the Boltzmann equation based on the isotropic transport
times. For $E>0$, the Boltzmann equation includes both direct intraband and
interband elastic scattering; while for $E_{-}\left(  k_{R}\right)  <E<0$,
only intraband scattering in the lower band occurs. Due to the band valley
structure below the band crossing point, the solution in this regime is
nontrivial and completely different from ordinary single-band cases. Finally
we clearly show that for positive and negative energies, the DFs are formally similar.

\subsection{The exact solution of the Boltzmann equation for $E>0$}

When $E>0$, the Boltzmann equation can be re-expressed as%
\begin{eqnarray}
& \mathbf{F}_{E}\cdot\mathbf{v}\left(  E,\lambda,\phi\right)  \partial
_{E}f^{0}=-\sum_{\lambda^{\prime}}N_{\lambda^{\prime}}\left(  E\right)
\int\frac{d\phi^{\prime}}{2\pi}\omega_{\lambda^{\prime},\lambda}^{\phi
^{\prime},\phi}\left(  E\right)  \nonumber\\
\times & \left[  g_{\lambda}\left(  E,\vartheta\left(  \mathbf{v}\left(
E,\lambda,\phi\right)  \right)  \right)  -g_{\lambda^{\prime}}\left(
E,\vartheta\left(  \mathbf{v}\left(  E,\lambda^{\prime},\phi^{\prime}\right)
\right)  \right)  \right]  ,\label{SBE2}%
\end{eqnarray}
where $\vartheta\left(  \mathbf{v}\right)  $ denotes the angle of the
direction of $\mathbf{v}$ with respect to that of the applied generalized
force $\mathbf{F}_{E}=-\frac{E-\mu}{T}\mathbf{\nabla}T-\mathbf{\nabla}%
\mu+e\mathbf{E}$. Due to Eq. (\ref{velocity1}), $\vartheta\left(
\mathbf{v}\left(  E,\lambda,\phi\right)  \right)  =\vartheta\left(
\mathbf{k}_{\lambda}\left(  E\right)  \right)  $. Above Boltzmann equation can be solved by
introducing the isotropic transport time for electrons with energy $E$ in the
$\lambda$ band as%
\begin{equation}
g_{\lambda}\left(  E,\vartheta\left(  \mathbf{k}_{\lambda}\left(  E\right)
\right)  \right)  =\left(  -\partial_{E}f^{0}\right)  \mathbf{F}_{E}%
\cdot\mathbf{v}\left(  E,\lambda,\phi\right)  \tau_{\lambda}\left(  E\right)
,\label{DF>1}%
\end{equation}
and the transport time is determined self-consistently by substituting Eq.
(\ref{DF>1}) into Eq. (\ref{SBE2}). Thus we obtain \
\begin{equation}
\frac{1}{\tau_{\lambda}\left(  E\right)  }=\sum_{\lambda^{\prime}}%
N_{\lambda^{\prime}}\left(  E\right)  \int\frac{d\phi^{\prime}}{2\pi}%
\omega_{\lambda^{\prime},\lambda}^{\phi^{\prime},\phi}\left(  E\right)
\left[  1-\cos\left(  \phi^{\prime}-\phi\right)  \frac{\tau_{\lambda^{\prime}%
}\left(  E\right)  }{\tau_{\lambda}\left(  E\right)  }\right]  ,\label{solve>}%
\end{equation}
where we use the relation $\left\vert \mathbf{v}\left(  E,\lambda^{\prime
},\phi^{\prime}\right)  \right\vert =\left\vert \mathbf{v}\left(
E,\lambda,\phi\right)  \right\vert $\ suitable for Rashba 2DES. In fact, Eq.
(\ref{solve>}) contains two coupled linear equations determining $\tau_{+}$
and $\tau_{-}$, yields
\begin{equation}
\tau_{\lambda}\left(  E\right)  =\tau_{0}\frac{N_{\lambda}\left(  E\right)
}{N_{0}}\left(  \frac{2N_{0}}{N_{>}\left(  E\right)  }\right)  ^{2}%
.\label{transport time>}%
\end{equation}
where $N_{>}\left(  E\right)  =2N_{0}$.
Combining with Eq. (\ref{velocity1}), the out-of-equilibrium DF takes a compact
form%
\begin{equation}
g_{\lambda}\left(  E,\vartheta\left(  \mathbf{k}_{\lambda}\left(  E\right)
\right)  \right)  =\left(  -\partial_{E}f^{0}\right)  \mathbf{F}_{E}\cdot
\frac{\hbar\mathbf{k}_{\lambda}\left(  E\right)  }{m}\tau_{0},\label{DF>2}%
\end{equation}
which satisfies the particle number conservation requirement%
\begin{equation}
\sum_{\lambda}\int dEN_{\lambda}\left(  E\right)  \int\frac{d\phi}{2\pi
}g_{\lambda}\left(  E,\vartheta\left(  \mathbf{k}_{\lambda}\left(  E\right)
\right)  \right)  =0.
\end{equation}
Eq. (\ref{DF>2}) looks similar to the DF for spin degenerate free electron gas
without SOC\cite{Ziman1972}, but the important difference is that in the
present case the group velocity is given by Eq. (\ref{velocity1}) rather than
$\frac{\hbar\mathbf{k}_{\lambda}\left(  E\right)  }{m}$, for positive
energies. This solution is the same as that obtained by employing a custom-designed
ansatz for the DF\cite{Trushin2007} in anisotropic
Rashba-Dresselhaus 2DES, while our approach is based on the simple physical
picture of isotropic transport time on constant-energy circles. On the other hand, this
transport time solution is different from the modified relaxation time
approximation (MRTA) solution for isotropic multiband systems: $g_{l}%
^{MRTA}=\left(  -\frac{\partial f^{0}}{\partial E_{l}}\right)  \mathbf{F}%
_{l}\cdot\mathbf{v}_{l}\tau_{l}^{MRTA}$ where
\begin{equation}
\frac{1}{\tau_{l}^{MRTA}}=\sum_{l\prime}\omega_{l\prime,l}\left[
1-\frac{\left\vert \mathbf{v}_{l\prime}\right\vert }{\left\vert \mathbf{v}%
_{l}\right\vert }\cos\left(  \vartheta\left(  \mathbf{v}_{l}\right)
-\vartheta\left(  \mathbf{v}_{l\prime}\right)  \right)  \right]  .\label{RTA}%
\end{equation}
Comparing Eq. (\ref{RTA}) with (\ref{solve>}), it is obvious that this MRTA
solution can not be self-consistently obtained from the Boltzmann equation for
Rashba 2DEG, because if one substitutes above $g_{l}^{MRTA}$ into the Boltzmann equation, it
is Eq. (\ref{solve>}) rather than Eq. (\ref{RTA}) that will be arrived at for
$\tau_{l}^{MRTA}$. Thus the MRTA only makes sense as an approximate solution
for Rashba 2DES. We can further point out that, in the present model, the
result of Eq. (\ref{RTA}) is the same as Eq. (\ref{transport time>}) only in the
zeroth order of SOC while different from the latter even in the 1st order of
SOC in the small SOC limit. Thus this MRTA in Rashba 2DEG even can not be
regarded as a better one than the constant RTA obtained by neglecting the
scattering-in term directly.

\subsection{The exact solution of the Boltzmann equation for $E_{-}\left(  k_{R}\right)
<E<0$}

When $E_{-}\left(  k_{R}\right)  <E<0$, by
converting the momentum integration in Eq. (\ref{SBE}) into energy integration
and noticing the different orientations of group velocity in the two monotonic branches,
the Boltzmann equation can be re-expressed as%
\begin{eqnarray}
& \mathbf{F}_{E}\cdot\mathbf{v}\left(  E,-\nu,\phi\right)  \partial_{E}%
f^{0}=-\sum_{\nu^{\prime}}N_{-\nu^{\prime}}\left(  E\right)  \int\frac
{d\phi^{\prime}}{2\pi}\omega_{-\nu^{\prime},-\nu}^{\phi^{\prime},\phi}\left(
E\right)  \nonumber\\
& \times\left[  g_{-\nu}\left(  E,\vartheta\left(  \mathbf{v}\left(
E,-\nu,\phi\right)  \right)  \right)  -g_{-\nu^{\prime}}\left(  E,\vartheta
\left(  \mathbf{v}\left(  E,-\nu^{\prime},\phi^{\prime}\right)  \right)
\right)  \right]  ,\label{SBE3}%
\end{eqnarray}
which is similar to Eq. (\ref{SBE2}) for $E>0$.
The derivation of the transport time solution of Eq. (\ref{SBE3}) is thus similar
to that when $E>0$. Substituting
\begin{equation}
g_{-\nu}\left(  E,\vartheta\left(  \mathbf{v}\left(  E,-\nu,\phi\right)
\right)  \right)  =\left(  -\partial_{E}f^{0}\right)  \mathbf{F}_{E}%
\cdot\mathbf{v}\left(  E,-\nu,\phi\right)  \tau_{-\nu}\left(  E\right)
\label{DF<1}%
\end{equation}
into Eq. (\ref{SBE3}), taking into account the fact that in the band valley
regime the direction of the group velocity can be parallel or anti-parallel to
that of the momentum, i.e., $\vartheta\left(  \mathbf{v}\left(  E,-1,\phi
\right)  \right)  =\vartheta\left(  \mathbf{k}_{-1}\left(  E\right)  \right)
$, $\vartheta\left(  \mathbf{v}\left(  E,-2,\phi\right)  \right)
=\vartheta\left(  \mathbf{k}_{-2}\left(  E\right)  \right)  +\pi$ and then
\begin{equation}
\frac{\cos\vartheta\left(  \mathbf{v}\left(  E,-\nu^{\prime},\phi^{\prime
}\right)  \right)  }{\cos\vartheta\left(  \mathbf{v}\left(  E,-\nu
,\phi\right)  \right)  }=\left(  -1\right)  ^{\nu^{\prime}-\nu}\frac
{\cos\vartheta\left(  \mathbf{k}_{-\nu^{\prime}}\left(  E\right)  \right)
}{\cos\vartheta\left(  \mathbf{k}_{-\nu}\left(  E\right)  \right)
},\label{direction}%
\end{equation}
we get the following self-consistent equation for $\tau_{-\nu}$:
\begin{eqnarray}
\frac{1}{\tau_{-\nu}\left(  E\right)  }  & =\sum_{\nu^{\prime}}N_{-\nu
^{\prime}}\left(  E\right)  \int\frac{d\phi^{\prime}}{2\pi}\omega
_{-\nu^{\prime},-\nu}^{\phi^{\prime},\phi}\left(  E\right)  \nonumber\\
& \times\left[  1-\left(  -1\right)  ^{\nu^{\prime}-\nu}\cos\left(
\phi^{\prime}-\phi\right)  \frac{\tau_{-\nu^{\prime}}\left(  E\right)  }%
{\tau_{-\nu}\left(  E\right)  }\right]  .\label{solve<}%
\end{eqnarray}
Here we have used the relation $\left\vert \mathbf{v}\left(  E,-\nu^{\prime
},\phi^{\prime}\right)  \right\vert =\left\vert \mathbf{v}\left(  E,-\nu
,\phi\right)  \right\vert $. Then the transport time is found as%
\begin{equation}
\tau_{-\nu}\left(  E\right)  =\tau_{0}\frac{N_{-\nu}\left(  E\right)  }{N_{0}%
}\left(  \frac{2N_{0}}{N_{<}\left(  E\right)  }\right)  ^{2}%
.\label{transport time<}%
\end{equation}
where $\left(  \frac{2N_{0}}{N_{<}\left(  E\right)  }\right)  ^{2}=\frac
{E_{R}^{2}+2E_{R}E}{E_{R}^{2}}$. Comparing this transport time for negative
energies to Eq. (\ref{transport time>}) for positive energies, one can see
that they share the same form.\

Therefore the nonequilibrium DF satisfying the particle number conservation
requirement is%
\begin{equation}
g_{-\nu}\left(  E\right)  =\left(  -\partial_{E}f^{0}\right)  \mathbf{F}%
_{E}\cdot\left[  \left(  -1\right)  ^{\nu-1}\frac{\hbar\mathbf{k}_{-\nu
}\left(  E\right)  }{m}\right]  \tau_{0}\left(  \frac{2N_{0}}{N_{<}\left(
E\right)  }\right)  ^{2}. \label{DF<2}%
\end{equation}
Here and below we use the simplified notation $g_{\lambda}\left(  E\right)  $
and $g_{-\nu}\left(  E\right)  $ to represent the DF for brevity. It is
obvious that this DF for negative energies is formally similar to that for
positive enegies when the latter is re-expressed as%
\begin{equation}
g_{\lambda}\left(  E\right)  =\left(  -\partial_{E}f^{0}\right)
\mathbf{F}_{E}\cdot\frac{\hbar\mathbf{k}_{\lambda}\left(  E\right)  }{m}%
\tau_{0}\left(  \frac{2N_{0}}{N_{>}\left(  E\right)  }\right)  ^{2},
\end{equation}
except one significant difference: the $\left(  -1\right)  ^{\nu-1}$ factor
for negative energies. This factor denotes nothing but the important fact
that, for electrons with negative energy on the branch $\nu=2$, the group
velocity and momentum has the opposite directions (see Eq. (\ref{velocity2})
and Fig.1).

By Eqs. (\ref{DF>2}) and (\ref{DF<2}), the out-of-equilibrium DFs in above two
energy regimes are continuous at $E=0$: $g_{+}\left(  E\rightarrow
0^{+}\right)  =g_{-2}\left(  E\rightarrow0^{-}\right)  =0$, $g_{-1}\left(
E\rightarrow0^{-}\right)  =g_{-}\left(  E\rightarrow0^{+}\right)  $.

\section{Electric-field and temperature-gradient induced spin polarization}

Since the nonequilibrium state is driven by the effective electric field
$\mathbf{E}^{\ast}=\mathbf{E}-\frac{1}{e}\mathbf{\nabla}\mu$ and temperature
gradient $\left(  -\mathbf{\nabla}T\right)  $, the spin density response takes
the following form in the linear response regime
\begin{equation}
\left\langle \mathbf{\hat{\sigma}}\right\rangle _{ext}=\boldsymbol{\chi
}_{\mathbf{E}}\cdot\mathbf{E}^{\ast}+\boldsymbol{\chi}_{{\small \nabla T}%
}\cdot\left(  -\mathbf{\nabla}T\right)  .
\end{equation}
Here the EISP coefficient $\boldsymbol{\chi}_{\mathbf{E}}$ and TISP
coefficient $\boldsymbol{\chi}_{{\small \nabla T}}$ can be calculated\textbf{
}from the 2nd equation of Eq. (\ref{SP1}), where the momentum integration is
performed by integrating over energy and polar angle. Substituting the out of
equilibrium DF, i.e., Eqs. (\ref{DF>2}) and (\ref{DF<2}), the
$\boldsymbol{\chi}_{\mathbf{E}}$ and $\boldsymbol{\chi}_{{\small \nabla T}}$
containing contributions from both bands\ are given by $\boldsymbol{\chi
}_{\mathbf{E}}\left(  T,\mu\right)  =\boldsymbol{\chi}_{\mathbf{E,+}}\left(
T,\mu\right)  +\boldsymbol{\chi}_{\mathbf{E,-}}\left(  T,\mu\right)  $:%
\begin{eqnarray}
\boldsymbol{\chi}_{\mathbf{E,+}}\left(  T,\mu\right)  =e\int\frac{d\phi}{2\pi
}\int_{0}^{\infty}dEN_{+}\left(  E\right)  \left(  -\partial_{E}f^{0}\right)
\mathbf{v}\left(  E,+,\phi\right)  \nonumber\\
\times\tau_{+}\left(  E\right)  \langle u_{\mathbf{k}_{+}\left(  E\right)
}|\mathbf{\sigma}|u_{\mathbf{k}_{+}\left(  E\right)  }\rangle,\nonumber\\
\boldsymbol{\chi}_{\mathbf{E,-}}\left(  T,\mu\right)  =e\int\frac{d\phi}{2\pi
}\int_{0}^{\infty}dEN_{-}\left(  E\right)  \left(  -\partial_{E}f^{0}\right)
\mathbf{v}\left(  E,-,\phi\right)  \nonumber\\
\times\tau_{-}\left(  E\right)  \langle u_{\mathbf{k}_{-}\left(  E\right)
}|\mathbf{\sigma}|u_{\mathbf{k}_{-}\left(  E\right)  }\rangle\nonumber\\
+e\int\frac{d\phi}{2\pi}\int_{E_{-}\left(  k_{R}\right)  }^{0}dE\sum_{\nu
}N_{-\nu}\left(  E\right)  \left(  -\partial_{E}f^{0}\right)  \label{EISP'}\\
\times\mathbf{v}\left(  E,-\nu,\phi\right)  \tau_{-\nu}\left(  E\right)
\langle u_{\mathbf{k}_{-\nu}\left(  E\right)  }|\mathbf{\sigma}|u_{\mathbf{k}%
_{-\nu}\left(  E\right)  }\rangle,\nonumber
\end{eqnarray}
and%
\begin{equation}
\boldsymbol{\chi}_{{\small \nabla T}}\left(  T,\mu\right)  =\frac{1}{e}%
\int_{E_{-}\left(  k_{R}\right)  }^{\infty}dE\left(  -\partial_{E}%
f^{0}\right)  \frac{E-\mu}{T}\boldsymbol{\chi}_{\mathbf{E}}\left(  E\right)
.\label{GE}%
\end{equation}
Here and below we use the simplified notation $\boldsymbol{\chi}_{\mathbf{E}%
}\left(  E\right)  $ to represent the zero-temperature EISP coefficient
$\boldsymbol{\chi}_{\mathbf{E}}\left(  T=0,E\right)  $ for brevity.
$\boldsymbol{\chi}_{\mathbf{E}}$ is a tensor and has two indices:
$\boldsymbol{\chi}_{\mathbf{E}}\left(  i,j\right)  $ where $i$ specifies the
spin component, and $j$ the direction of the electric field. Due to the
isotropy, we can apply the generalized force only in $x$ direction for the
calculation, and only $\boldsymbol{\chi}_{\mathbf{E}}\left(  \hat{y},\hat
{x}\right)  $ will be calculated below ($\boldsymbol{\chi}_{\mathbf{E}}\left(
\hat{x},\hat{x}\right)  =0$). Also we will drop the indices $\left(  \hat
{y},\hat{x}\right)  $ in $\boldsymbol{\chi}_{\mathbf{E}}\left(  \hat{y}%
,\hat{x}\right)  $, for simplicity, $\chi_{\mathbf{E}}$.

\subsection{EISP}

The zero-temperature EISP for $E_{F}\geq0$ can be obtained easily from Eq.
(\ref{EISP'}) as
\begin{equation}
\chi_{\mathbf{E}}\left(  E_{F}\geq0\right)  =e\tau_{0}\frac{\alpha}{\hbar
}2N_{0}\mathbf{.} \label{Edelstein}%
\end{equation}
This result has been well-known since Edelstein\cite{Edelstein1990}. It is
independent on the Fermi energy, because the directions of spin on the inner
$\left(  +\right)  $ and outer $\left(  -\right)  $ Fermi circles are opposite
at the same polar angle $\phi$ and the $E_{F}$-dependence of EISPs in both
Fermi circles cancels.

While, for Fermi energies below the band crossing point, the EISP takes the
following non-Edelstein form%
\begin{equation}
\chi_{\mathbf{E}}\left(  E_{F}\leq0\right)  =e\tau_{0}\frac{\alpha}{\hbar
}2N_{0}\left(  1+\frac{2E_{F}}{E_{R}}\right)  \mathbf{.} \label{non-Edelstein}%
\end{equation}
It is linearly dependent on the Fermi energy, different from the positive
Fermi energy case. In the band valley, the orientations of spin on the inner
$\left(  -2\right)  $ and outer $\left(  -1\right)  $ Fermi circles are
parallel at the same $\phi$ and the $E_{F}$-dependence of EISPs of both Fermi
circles does not cancel. Since the Fermi surface topology in the band valley
differs from that above the band crossing point, the behaviors of EISPs are
different between the two regimes. $\chi_{\mathbf{E}}\left(  E\right)  $ is
continuous at $E=0$. The contribution to $\chi_{\mathbf{E}}\left(  0\right)  $
entirely comes from the outer Fermi circle $\left(  E_{F}=0,k=2k_{R}\right)
$, since the DOS at the band crossing point $\left(  E_{F}=0,k=0\right)  $ is zero.

We compare Eqs. (\ref{Edelstein}) and (\ref{non-Edelstein}) to the EISP
obtained by employing the constant RTA \cite{Dyrdal2013}:
$\chi_{\mathbf{E}}^{RTA}\left(  E_{F}\geq0\right)  =e\tau\frac{\alpha}{\hbar
}2N_{0}$, $\chi_{\mathbf{E}}^{RTA}\left(  E_{F}\leq0\right)  =e\tau
\frac{\alpha}{\hbar}2N_{0}\sqrt{1+\frac{2E_{F}}{E_{R}}}$, with $\tau$ the
constant relaxation time independent on the energy and band. For $E_{F}\geq0$
the EISP obtained by the constant RTA has the same Edelstein form, while when
$E_{F}\leq0$ the constant RTA result shows different $E_{F}$-dependence\ from
Eq. (\ref{non-Edelstein}).

\begin{figure}[ptb]
\begin{indented}
\item[]\includegraphics[width=0.35\textwidth]{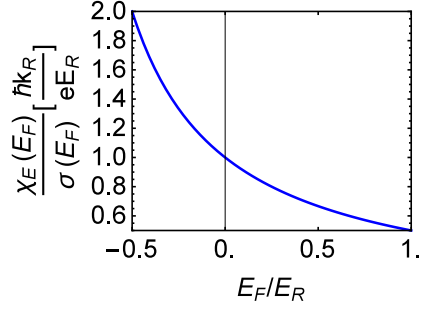}
\end{indented}
\caption{Spin polarization efficiency.}%
\label{fig2}%
\end{figure}Now we calculate the spin polarization efficiency, defined as the
ratio between the electrical-field induced spin polarization and the driven
electric-current density. The electrical conductivity for the same model has
been given by \cite{Xiao2}: $\sigma\left(  E_{F}\geq0\right)  =\frac{e^{2}%
}{2\pi^{2}\hbar}\frac{2\pi\left(  E_{F}+E_{R}\right)  \tau_{0}}{\hbar}$,
$\sigma\left(  E_{F}\leq0\right)  =\frac{e^{2}}{2\pi^{2}\hbar}\frac
{2\pi\left(  E_{F}+E_{R}\right)  \tau_{0}}{\hbar}\left(  1+\frac{2E_{F}}%
{E_{R}}\right)  $. Therefore, the spin polarization efficiency is given by
a single expression suitable for both positive and negative Fermi energies%
\begin{equation}
\frac{\chi_{\mathbf{E}}\left(  E_{F}\right)  }{\sigma\left(  E_{F}\right)
}=\frac{\hbar k_{R}}{eE_{R}}\frac{1}{1+\frac{E_{F}}{E_{R}}},
\label{efficiency}%
\end{equation}
which increases for decreased Fermi energies (shown in Fig. 2). Eqs.
(\ref{Edelstein}), (\ref{non-Edelstein}) and (\ref{efficiency}) show that when
the Fermi energy lies below the band crossing point, alough the EISP is
lowered, higher spin polarization efficiency is achieved.

\subsection{TISP}

We substitute $\chi_{\mathbf{E}}\left(  E\right)  $ into Eq. (\ref{GE}), and
define%
\[
\frac{E-\mu}{k_{B}T}=x,\frac{\mu}{k_{B}T}=-t_{1},\frac{E_{F}}{k_{B}T}=-t_{2},
\]%
\begin{equation}
a\left(  t_{1}\right)  =\int_{t_{1}}^{\infty}dx\left(  -\frac{\partial f^{0}%
}{\partial x}\right)  x,b\left(  t_{1}\right)  =\int_{t_{1}}^{\infty}dx\left(
-\frac{\partial f^{0}}{\partial x}\right)  x^{2},
\end{equation}
then the TISP is found as%
\begin{equation}
\chi_{{\small \nabla T}}=\frac{k_{B}}{e}\chi_{\mathbf{E}}\left(  0\right)
\frac{2k_{B}T}{E_{R}}\frac{\pi^{2}}{3}\left[  1-\frac{b\left(  t_{1}\right)
-t_{1}a\left(  t_{1}\right)  }{\pi^{2}/3}\right]  .\label{main result1}%
\end{equation}

The relation between $t_{2}$ and $t_{1}$ (i.e., the chemical potential at low
temperatures) can be obtained by the consideration about electron
density\cite{Ziman1972}%
\begin{eqnarray}
t_{2}-t_{1}  &  =0,E_{F}\gg k_{B}T,\nonumber\\
t_{2}-t_{1}  &  =o\left(  \frac{k_{B}T}{E_{R}}\right)  ,\left\vert
E_{F}\right\vert \sim k_{B}T,\\
t_{2}-t_{1}  &  =\frac{\pi^{2}}{6}\frac{k_{B}T}{E_{R}+2E_{F}},-E_{F}\gg
k_{B}T.\nonumber
\end{eqnarray}
Here we only consider small thermal fluctuations $k_{B}T\ll E_{R}$ and
$E_{F}+\frac{1}{2}E_{R}\gg k_{B}T$, so that the band valley structure and the
Fermi surfaces survive the thermal smearing. In some new materials with giant
Rashba effect, e.g., strongly spin-orbit coupled 2DES near the surface of
Rashba semiconductors BiTeX (X=Cl, Br, I), $E_{R}$ is about $35\sim200meV$.
Therefore, $k_{B}T\ll E_{R}$ and $E_{F}+\frac{1}{2}E_{R}\gg k_{B}T$\ can be
satisfied at low temperatures about several Kelvins for not too low Fermi energies.

Consequently we can set $t_{2}=t_{1}$ in the expression for $\chi
_{{\small \nabla T}}$ since Eq. (\ref{main result1}) has already been
$o\left(  \frac{k_{B}T}{E_{R}}\right)  $:%
\begin{equation}
\chi_{{\small \nabla T}}=\tau_{0}\frac{\alpha}{\hbar}2N_{0}k_{B}\frac{\pi^{2}%
}{3}\left[  1-3\frac{b\left(  t_{2}\right)  -t_{2}a\left(  t_{2}\right)  }%
{\pi^{2}}\right]  \frac{2k_{B}T}{E_{R}}\mathbf{.} \label{main result}%
\end{equation}
This is our main analytical result for TISP. This formula is still valid for Fermi energies
near the band crossing point, where the Sommerfeld
expansion is not suitable for the treatment of Eq. (\ref{GE}), due to the fact
that the energy-derivative of $\boldsymbol{\chi}_{\mathbf{E}}\left(E\right)$ is not continuous
at the band crossing point.

\begin{figure}[ptb]
\begin{indented}
\item[]\includegraphics[width=0.4\textwidth]{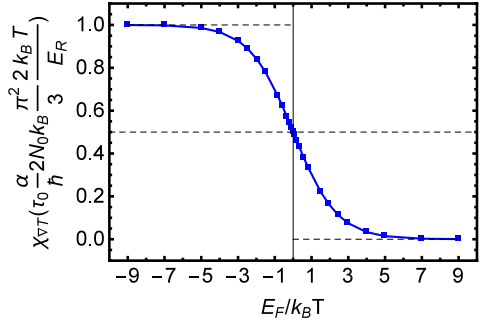}
\end{indented}
\caption{Temperature-gradient induced spin polarization.}%
\label{fig3}%
\end{figure}According to Eq. (\ref{main result}), $\chi_{\mathbf{\nabla}T}$ is
given in Fig. 3, in units of $\tau_{0}\frac{\alpha}{\hbar}2N_{0}k_{B}\frac{\pi
^{2}}{3}\frac{2k_{B}T}{E_{R}}$. When $E_{F}/k_{B}T\gtrsim5$, $\chi
_{\mathbf{\nabla}T}$ almost vanishes. When $E_{F}/k_{B}T\lesssim-5$,
$\chi_{\mathbf{\nabla}T}/\tau_{0}\frac{\alpha}{\hbar}2N_{0}k_{B}\frac{\pi^{2}}%
{3}\frac{2k_{B}T}{E_{R}}$ approaches a constant value $1$. In the intermediate
regime $-5\lesssim E_{F}/k_{B}T\lesssim5$, $\chi_{\mathbf{\nabla}T}$ is
monotonically decreasing as $E_{F}/k_{B}T$ increases.

Eq. (\ref{main result}) can be re-arranged as
\begin{equation}
\chi_{{\small \nabla T}}=\tau_{0}\frac{\alpha}{\hbar}2N_{0}k_{B}\frac{\pi^{2}%
}{3}\frac{2\left\vert E_{F}\right\vert }{E_{R}}\left[  1-\frac{b\left(
t_{2}\right)  -t_{2}a\left(  t_{2}\right)  }{\pi^{2}/3}\right]  \frac
{1}{\left\vert -t_{2}\right\vert }\mathbf{.}%
\end{equation}
To make the temperature dependence of $\chi_{\mathbf{\nabla}T}$ clear, we
replot $\chi_{\mathbf{\nabla}T}$ (in units of $\tau_{0}\frac{\alpha}{\hbar
}2N_{0}k_{B}\frac{\pi^{2}}{3}\frac{2\left\vert E_{F}\right\vert }{E_{R}}$) as
function of $\frac{k_{B}T}{\left\vert E_{F}\right\vert }$ ($\frac
{1}{\left\vert -t_{2}\right\vert }$) in Fig. 4. \begin{figure}[ptb]
\begin{indented}
\item[]\includegraphics[width=0.5\textwidth]{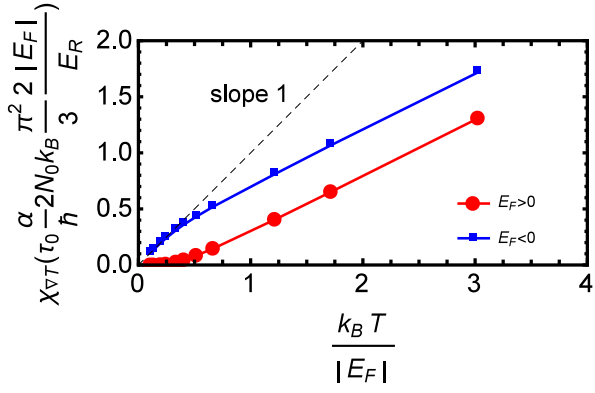}
\end{indented}
\caption{The Temperature dependence of\textbf{ }$\chi_{\mathbf{\nabla}T}$ for
positive Fermi energy (red curve) and negative Fermi energy (blue curve).}%
\label{fig4}%
\end{figure}

It shows that $\chi_{\mathbf{\nabla}T}$ increases with increasing temperature.
When $k_{B}T/E_{F}\lesssim0.2$, $\chi_{\mathbf{\nabla}T}$ fully vanishes, when
$k_{B}T/E_{F}$ continues to increase the contribution of the band valley
structure will be included and dominates $\chi_{\mathbf{\nabla}T}$, then
$\chi_{\mathbf{\nabla}T}$ is enhanced. When $k_{B}T/\left(  -E_{F}\right)
\lesssim0.2$, $\chi_{{\small \nabla T}}/\tau_{0}\frac{\alpha}{\hbar}2N_{0}k_{B}%
\frac{\pi^{2}}{3}\frac{2\left(  -E_{F}\right)  }{E_{R}}$ is almost exactly
linear in $k_{B}T/\left(  -E_{F}\right)  $ with slope $1$. When the
temperature increases, not only the band valley contributes, but also the
electron states above the band crossing point are included due to the thermal
smearing, thus the TISP is suppressed. Moreover, when $\frac{k_{B}%
T}{\left\vert E_{F}\right\vert }\gg1$, $t_{2}\rightarrow0$, $1-3\frac{b\left(
t_{2}\right)  -t_{2}a\left(  t_{2}\right)  }{\pi^{2}}\rightarrow\frac{1}{2}$,
so $\chi_{{\small \nabla T}}\left(  \frac{k_{B}T}{\left\vert E_{F}\right\vert
}\gg1\right)  \rightarrow\chi_{{\small \nabla T}}\left(  E_{F}=0\right)
=\tau_{0}\frac{\alpha}{\hbar}2N_{0}k_{B}\frac{\pi^{2}}{3}\frac{1}{2}%
\frac{2k_{B}T}{E_{R}}$ linear in $T$.

\section{Conclusions}

In conclusion, we have calculated the thermoelectric responses of spin
polarization in 2D Rashba system. By self-consistently determining the
transport time, we exactly solved the Boltzmann equation when static impurity
scatterings dominate the electron relaxation process. It was shown that the
electric-field induced spin polarization is linearly dependent on the Fermi
energy when only the lower band is occupied, different from the Edelstein
behavior when both bands occupied. Higher spin polarization efficiency is
achieved when the Fermi energy lies below the band crossing point. It was
found that the temperature-gradient induced spin polarization continuously
increase to a saturation value as the Fermi energy decreases below the band
crossing point. In addition, the temperature-gradient induced spin
polarization tends to zero at vanishing temperatures.

This work may stimulate more experimental and theoretical works on the
electrical and thermal spin control in Rashba semiconductors BiTeX (X=Cl, Br,
I) and BiTeX quantum wells, as well as other materials with giant Rashba spin splitting.

\ack
The Work is supported by National Natural Science Foundation of China (No.
11274013 and No. 11274018), and Ministry of Science and Technology of the People's Republic of China (2012CB921300).

\end{document}